\documentclass[11pt,a4paper]{article}

\usepackage[latin1]{inputenc}
\usepackage[english]{babel}
\usepackage{amsmath,amssymb}
\usepackage{graphicx}
\usepackage{cite}
\usepackage{place ins}

\newcommand{\as}{\alpha_s}

\newcommand{\beq}{\begin{equation}}
\newcommand{\eeq}{\end{equation}}
\newcommand{\bea}{\begin{eqnarray}}
\newcommand{\eea}{\end{eqnarray}}
\newcommand{\bdm}{\begin{displaymath}}
\newcommand{\edm}{\end{displaymath}}
\def\as{\alpha_s}

\def\ord{{\cal O}}

\def\dzero {D\O \;}

\begin{document}
\begin{titlepage}
\begin{flushright}
{DCPT/12/106 \\ IPPP/12/53}
\end{flushright}     

\begin{center}
{\Large \bf
Phenomenology of electro-weak bosons at hadron colliders with novel variables}
\vspace*{1.5cm}

 Simone Marzani
\\
\vspace{0.3cm}  {\it
Institute for Particle Physics Phenomenology, \\Durham University,\\ Durham DH1 3LE, United Kingdom}
\vspace*{1.5cm}

\bigskip
\bigskip

 { \bf Abstract }
\end{center}
\begin{quote}
We review the phenomenology of electro-weak bosons produced in hadron-hadron collisions. In particular, we discuss the transverse momentum distribution of lepton pairs with invariant mass close to the $Z$ peak.
We describe the theoretical calculation for the magnitude of the transverse momentum $Q_T$ and its comparison to Tevatron and LHC data. We also discuss the related variable $\phi^*$, describing its experimental advantages as well as its relation to the standard $Q_T$ variable. Finally, we compare resummed predictions for $\phi^*$ to data.
\end{quote}

\bigskip
\bigskip

\emph{e-mail}: simone.marzani@durham.ac.uk
\end{titlepage}

\section{Introduction} \label{sec:introduction}
The production of a lepton pair $l^+ l^-$ in hadron-hadron collisions is one of the most extensively studied processes in particle phenomenology, with the original paper appearing more than forty years ago~\cite{DY}.  Since then a huge theoretical effort has gone into improving the accuracy of the predictions. For instance, QCD corrections are known to next-to--next-to leading order (NNLO), not only for the inclusive rate~\cite{DYNNLO}, but also for rapidity distributions~\cite{DYNNLOrap} and more recently for fully exclusive cross-sections~\cite{FEWZ, FEWZ2,FEWZ2_1}. 

Of particular interest is the transverse momentum distribution of the lepton pair, or equivalently of the gauge boson decaying into it. At Born level, the production process is described by quark-antiquark annihilation diagrams.
In the framework of collinear factorisation, the initial-state (anti)quarks are on-shell and they move in same direction as the incoming hadrons. Thus, at this perturbative order the vector boson ($Z/\gamma^*$), and hence the lepton pair,  is produced with no transverse momentum with respect to the beam. The vector boson acquires a non-zero transverse momentum when we consider QCD radiative corrections. The $Z/\gamma^*$ transverse momentum distribution is therefore sensitive to multi-gluon emission from the initial-state partons, while having a particularly simple final state, and hence provides a powerful tool to test our understanding of QCD dynamics. 

The observable we are consideing is a typical example of a multi-scale problem and the correct treatment of multi-gluon emissions goes beyond fixed-order perturbation theory. Let us introduce the invariant mass of the leptons $M$, that will be chosen around the $Z$ mass, and the let us call $Q_T$ the magnitude of the vector boson transverse momentum.
We are going to consider three different regimes. When $Q_T\sim M$ we expect fixed order perturbation theory to work.
In the region $\Lambda_{{\rm QCD}}\ll Q_T\ll M$, we can still rely on perturbation theory but large logarithms of the ratio $Q_T/M$ may spoil the convergence of the perturbative expansion and must be resummed to all orders~\cite{DDT, APP, davies, CSS,ERV, EV,CdFGuniv, GrazzDeFlo,GrazziniDeFlorian, FlorenceQT, FlorenceDY}. Finally, in the region $Q_T\sim \Lambda_{{\rm QCD}}$ we expect non-perturbative effects to play a significant role. These contributions are associated with the intrinsic transverse momentum of the initial state partons, which is neglected in the usual perturbative treatment based on collinear factorisation.
Therefore, it is important to compute a solid perturbative prediction, so that we can compare it to precise data coming from the experiments and be able to eventually pin down non-perturbative contributions.

\section{The challenge of theoretical predictions at small $Q_T$} \label{sec:theory}

The leading order (LO) transverse momentum distribution can be obtained by integrating the $\ord(\as)$ corrections to the Born process~\cite{aurenche}. In the small-$Q_T$ limit we obtain the following behaviour
\beq \label{LOQT}
\frac{{\rm d} \sigma}{{\rm d} Q_T^2} = \frac{\alpha_s}{2 \pi}\left( \frac{\mathcal{A}}{Q_T^2} \ln \frac{M^2}{Q_T^2}+ \frac{\mathcal{B}}{Q_T^2}+ \mathcal{C}(Q_T^2)   \right) +\ord(\as^2). 
\eeq
The distribution above exhibits a double logarithmic divergence when $Q_T\to 0$, which corresponds to the emission of a soft and collinear gluon. The coefficient of the double logarithm (DL) is very simple: $\mathcal{A}= 2 C_F \Sigma_0$, where $\Sigma_0$ is the hadronic Born cross-section. The single-logarithmic term instead contains a contribution which originates from the emission of a hard collinear parton. For this reason the coefficient $\mathcal{B}$ does not trivially factorise and it contains a convolution term which depends on the parton distribution functions (PDFs).

The behaviour described at $\ord(\as)$ by Eq.~(\ref{LOQT}) is actually present to all orders in perturbation theory. At the DL level we have
\beq \label{DLQT}
\frac{{\rm d} \sigma}{{\rm d} Q_T^2} = \frac{\Sigma_0}{Q_T^2} \left[  \frac{\alpha_s}{2 \pi}A_1 \ln \frac{M^2}{Q_T^2}+ \left(\frac{\alpha_s}{2 \pi}\right)^2 A_2 \ln^3 \frac{M^2}{Q_T^2}   + \dots  + \left(\frac{\alpha_s}{2 \pi}\right)^n A_2 \ln^{2 n-1} \frac{M^2}{Q_T^2}+\dots \right]. 
\eeq
Not only is any truncation of the above series divergent in the formal limit $Q_T\to0$ but it also loses predictive power if $\as \ln \frac{M^2}{Q_T^2}=\ord(1)$, because all higher-orders are equally important. These contributions need to be resummed to all-orders in order to obtain a reliable answer.
The calculation in the soft and collinear limit is not difficult and we find that the coefficients $A_i$ are proportional to powers of the Casimir $C_F$. The DL series can be then summed to:
\beq \label{DLQTres}
\frac{d \sigma}{d Q_T^2} = \Sigma_0  \frac{d}{d Q_T^2} e^{ -\frac{\alpha_s}{2 \pi}C_F \ln^2 \frac{M^2}{Q_T^2} }.
\eeq
The resummed expression exhibits a Sudakov form factor that suppresses the emission of soft gluons. As a consequence, the distribution is now well-behaved and, actually, vanishes in the $Q_T\to 0$ limit. However, we may question the usefulness of Eq.~(\ref{DLQTres}) for phenomenology. In fact, the hypotheses used to derive the DL results are too crude to capture all the relevant effects,  though they are formally sub-leading. For instance, using momentum conservation in the transverse plane, we have
$
\underline{Q}_T= - \sum_i \underline{k}_{Ti},
$
where $\underline{k}_i$ are the transverse momenta of the emitted partons. It is then clear that small values of $Q_T=\sqrt{|\underline{Q}_T|^2}$ can be achieved via two distinct mechanisms: emission of gluons with small transverse momentum or kinematical cancellation between, not necessarily soft, $\underline{k}_{Ti}$. 

We would then like to extend the resummation formalism beyond the DL level. The derivation we sketch in the following is valid  at next-to-leading logarithmic accuracy (NLL), although the state of the art is actually one logarithmic order higher (NNLL). Firstly, we use the well-known property that convolutions of matrix-elements with PDFs can be written as ordinary products of their Mellin moments. Therefore, we are able to arrive at the distribution for the emission of an arbitrary number of collinear and optionally soft gluons in Mellin space:
\beq \label{almostfact}
\frac{{\rm d} \tilde{\sigma}}{{\rm d} Q_T^2}(N) =\tilde{\Sigma}_0(N) \sum_{n=0}^\infty \frac{1}{n!}\prod_{i=1}^n \int {\rm d} z_i  \frac{{\rm d} k_{Ti}}{k_{Ti}} \frac{{\rm d} \phi_{i}}{2 \pi} \frac{\as(k_{Ti})}{2 \pi} z_i^N 2 P_{qq}(z_i) \,\delta^{(2)}\left(\sum_{i=1}^n \underline{k}_{Ti}+\underline{Q}_T \right),
\eeq
where $P_{qq}$ is the Altarelli-Parisi splitting function and $\tilde{\Sigma}_0(N)$ is the Mellin transform of the hadronic Born cross-section
\beq \label{born}
\tilde{\Sigma}_0(N) =  \sigma_0 \sum_q f_{q/A}(N,\mu) f_{\bar{q}/B}(N,\mu).
\eeq

We note that the $\delta$ function that enforces momentum conservation in Eq.~(\ref{almostfact}) spoils the factorisation properties of the result.  This issue can be solved by introducing a two-dimensional impact-parameter
representation of the $\delta$ function
\beq \label{delta_rep}
\delta^{(2)}\left(\sum_{i=1}^n \underline{k}_{Ti}+\underline{Q}_T \right)=\frac{1}{4 \pi^2}\int {\rm d} ^2 \underline{b} \, e^{i\underline{b}\cdot \underline{Q}_T}\prod_{i=1}^n e^{i\underline{b}\cdot \underline{k}_{Ti}},
\eeq 
which has the desired factorised form. We can now perform the sum in Eq.~(\ref{almostfact}), obtaining an exponential:
\bea \label{fact}
\frac{{\rm d} \tilde{\sigma}}{{\rm d} Q_T^2}(N) &=&\tilde{\Sigma}_0(N)\frac{1}{4 \pi^2}\int {\rm d} ^2 \underline{b} \, \exp \left[ \int {\rm d} z_i  \frac{{\rm d} k_{T}}{k_{T}} \frac{{\rm d} \phi}{2 \pi} \frac{\as(k_{T})}{\pi}  (z^N e^{i\underline{b}\cdot \underline{k}_{T}}-1) P_{qq}(z) \right] \nonumber \\ &\equiv& \tilde{\Sigma}_0(N)\frac{1}{4 \pi^2}\int {\rm d} ^2 \underline{b} \, e^{-R_N(b,M,N)},
\eea
where the $(-1)$ term has been included to account for virtual corrections.
If we are interested in the $Q_T$ spectrum, we integrate over the angle between $\underline{b}$ and $\underline{Q}_T$, obtaining a Bessel function:
\beq \label{fact2}
\frac{{\rm d} \tilde{\sigma}}{{\rm d} Q_T^2}(N) =\frac{\tilde{\Sigma}_0(N)}{4 \pi^2}\int_0^\infty {\rm d} b b \int_0^{2 \pi}{\rm d} \varphi  \, e^{i b Q_T \cos \varphi} \, e^{-R_N(b,M,N)}=
 \frac{\tilde{\Sigma}_0(N)}{2 \pi}\int_0^{\infty} {\rm d} b b J_0(b Q_T) e^{-R_N(b,M,N)}.
\eeq

We can now have a closer look at the resummed exponent $R_N$; at NLL accuracy we find the following structure:
\beq \label{radN}
R_N(b,M,N)=  \int_{\bar{b}^{-2}}^{M^2} \frac{{\rm d} k_{T}^2}{k_{T}^2} \frac{\as(k_{T})}{\pi}  \left[  \left( A^{(1)}+ \frac{\as(k_{T})}{\pi}  A^{(2)} \right) \ln \frac{M^2}{k_{T}^2}+ B^{(1)}+2 \gamma_{qq}(N) \right],
\eeq
where $\bar{b}= b e^{\gamma_E}/2$ and we have introduced the coefficients:
\beq
A^{(1)}= C_F,\quad A^{(2)}= \frac{C_F}{2} \left[C_A \left(\frac{67}{18}-\frac{\pi^2}{6}\right)-\frac{5}{9}n_f\right], \quad B^{(1)}=-\frac{3}{2}C_F,
\eeq
and $\gamma_{qq}$ is the Mellin transform of the LO splitting function. The $N$ dependent part of the function $R_N$ can be absorbed into the PDFs using DGLAP evolution. The PDFs, which in Eq.~(\ref{almostfact}) are contained in the function $\tilde{\Sigma}_0$, are then evolved down to the scale $\mu=1/\bar{b}$.  In the following the part of $R_N$ which does not depend on $N$ will be denoted by $R$.

At NLL accuracy the integration over $k_T$ in Eq.~(\ref{radN}) must be performed with the two-loop expression for the running coupling. We can then express the resummed exponent as the sum of two contributions:
\begin{equation} \label{Rlambda}
 R(b) = L g^{(1)} (\alpha_s L ) + g^{(2)} (\alpha_s L)+\dots
\end{equation}
with
\begin{align}
  g^{(1)}(\lambda) =& \frac{A^{(1)}}{\pi\beta_0\lambda} \left
    [-\lambda-\ln{(1-\lambda)}\right]\,,\\
  g^{(2)}({\lambda})=&  \frac{-B^{(1)}}{\pi \beta_0} \ln(1-\lambda)
  +\frac{A^{(2)}[\lambda + (1-\lambda)\ln(1-\lambda)]}
  {\pi^2\beta_0^2(1-\lambda)} -\frac{A^{(1)}\beta_1}{\pi\beta_0^3}
  \left[ \frac{\lambda + \ln (1-\lambda)}{1-\lambda} + \frac{1}{2}
    \ln^2{(1-\lambda)} \right],
\label{eq:g2}
\end{align}
and $\lambda= \as(M) \beta_0 L$, $L=\ln \left(\bar{b}M\right)^2$. The function $g^{(1)}$ resums LL terms, while $g^{(2)}$ NLL ones.

As mentioned before, the state of the art in $Q_T$ resummation is actually NNLL. In order to achieve that accuracy the coefficients $B^{(2)}$~\cite{davies} and $A^{(3)}$~\cite{BecherNeubertA3} have been calculated. Moreover, the integrals in Eq.~(\ref{radN}) must be performed with the three-loop expression for the running coupling. This results in a new contribution, $ \as g^{(3)}$, to Eq.~(\ref{Rlambda}).

So far we have dealt with all the logarithmic terms in the function $R$. However, there is still one contribution we must consider in order to achieve (N)NLL accuracy:  non-logarithmic corrections to the Born cross-section $\tilde{\Sigma}_0$ in Eq.~(\ref{born}):
\bea
\tilde{\Sigma}&=& \sigma_0 \sum_{q} \sum_{\alpha,\beta} \left[ \delta_{q\alpha}+ \sum_{i=1}^n \left(\frac{\as(1/\bar{b})}{\pi} \right)^iC^{(i)}_{q \alpha}(N) \right] \left[ \delta_{\bar{q}\beta}+ \sum_{i=1}^n \left(\frac{\as(1/\bar{b})}{\pi} \right)^i C^{(i)}_{\bar{q} \beta}(N) \right] \nonumber \\ && f_{\alpha/A}(N,1/\bar{b}) f_{\beta/B}(N,1/\bar{b}),
\eea
in particular, the $i=1, 2$ coefficients are needed~\cite{EllisStirling, FlorenceDY}.

Lastly, in order to obtain a resummed expression for the $Q_T$ distribution, the integral over the impact parameter $b$ in Eq.~(\ref{fact}) must be performed. A closer look at Eq.~(\ref{Rlambda}) reveals that this integration is problematic both at small $b$ and large $b$, because the function $R$ diverges in both limits. The small-$b$ region corresponds to large values of $Q_T$, which are beyond the 
jurisdiction of the resummation formula.  Different prescriptions that preserve the logarithmic accuracy can be found in the literature. For instance, in Refs~\cite{FlorenceQT, FlorenceDY} the argument of the logarithms is modified: $\ln\left(\bar{b}^2M^2 \right) \to \ln\left(1+\bar{b}^2M^2 \right)$, so that $R(b)$ vanishes as $b\to0$. Alternatively~\cite{BDM,BDMT,BDMTlhc} we can freeze the radiator below some value $b_{\rm min}$.

The large-$b$ divergence, $\as \beta_0 L =1$ is due to the Landau pole in the running coupling and is associated to non-perturbative (NP) behaviour. In this case as well different prescriptions to deal with the singularity can be found in the literature. In Ref.~\cite{CSS} a smooth transition between a perturbative and NP region was derived in terms of functions that must be fitted from the data.  Improvements of this NP form factor were suggested in Refs.~\cite{resbos, resbos_comp}. Other approaches consist in introducing an upper limit to the $b$ integration~\cite{BDM,BDMT,BDMTlhc} or deforming the contour integration off the real axis, so to avoid the Landau singularity, as, for instance, in~\cite{FlorenceQT, FlorenceDY}.

Different, and equally successful, approaches to $Q_T$ resummation are the ones based on soft-collinear effective theory (SCET). Calculations have been performed to NNLL accuracy by two groups in~\cite{BecherNeubertA3,BecherNeubertPheno} and in~\cite{mantrypetriello1,mantrypetriello2,mantrypetriello3} and successfully compared to Tevatron and LHC data. 

\section{Comparison between theory and experiment} \label{QTcomparison}

The transverse momentum spectrum of vector bosons produced via the Drell-Yan mechanism has been extensively studied by the Tevatron experiments~\cite{CDFRunI,D0RunI, D0RunII} and, more recently, by the LHC collaborations as well~\cite{atlasZpt,cmsZpt}. State-of-the-art-perturbative predictions have been compared to data in order to test the accuracy of resummed calculations and to asses the role of NP effects.

\begin{figure}
\begin{center}
\includegraphics[width=0.495 \textwidth]{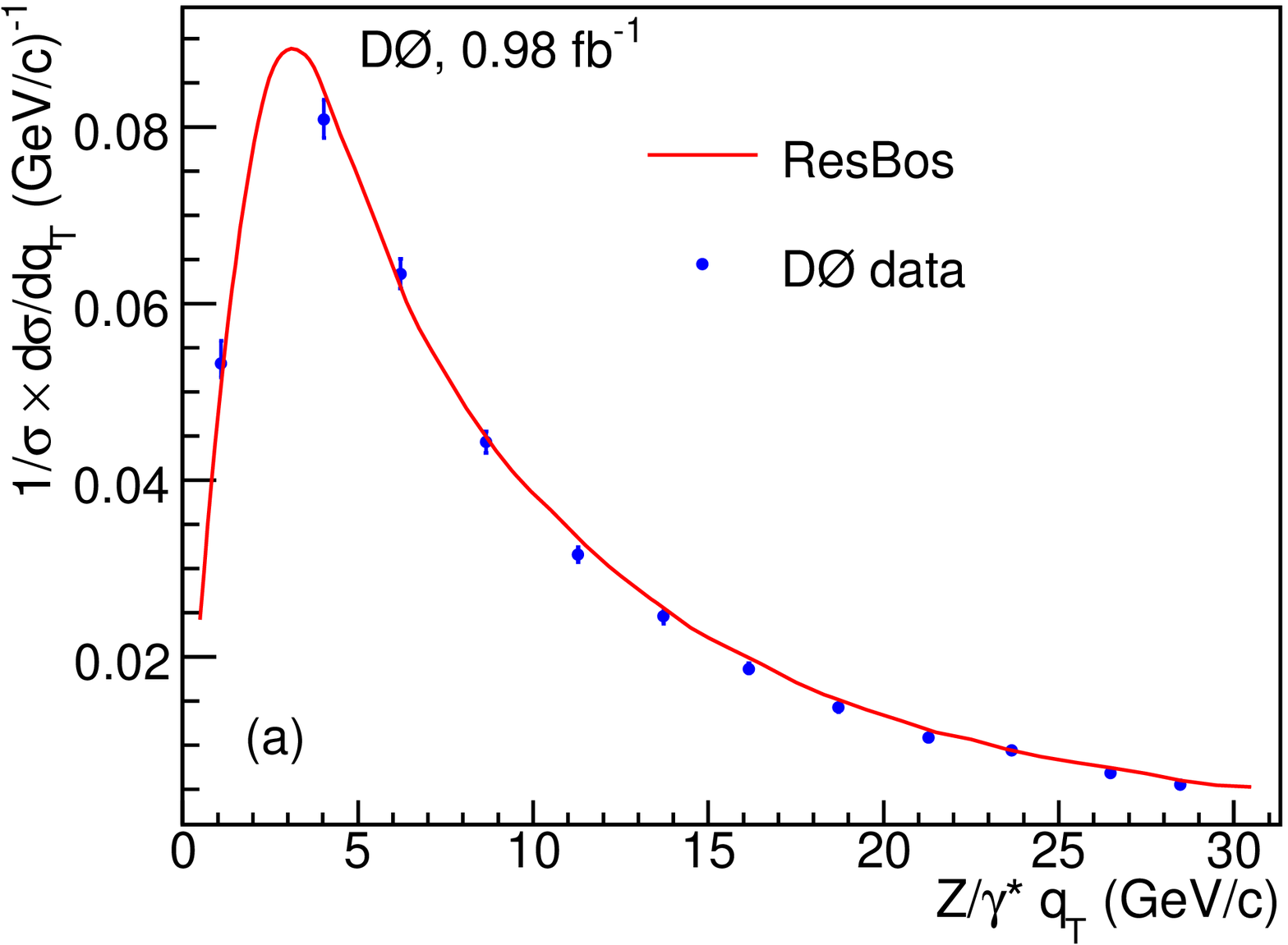}
\includegraphics[width=0.495 \textwidth]{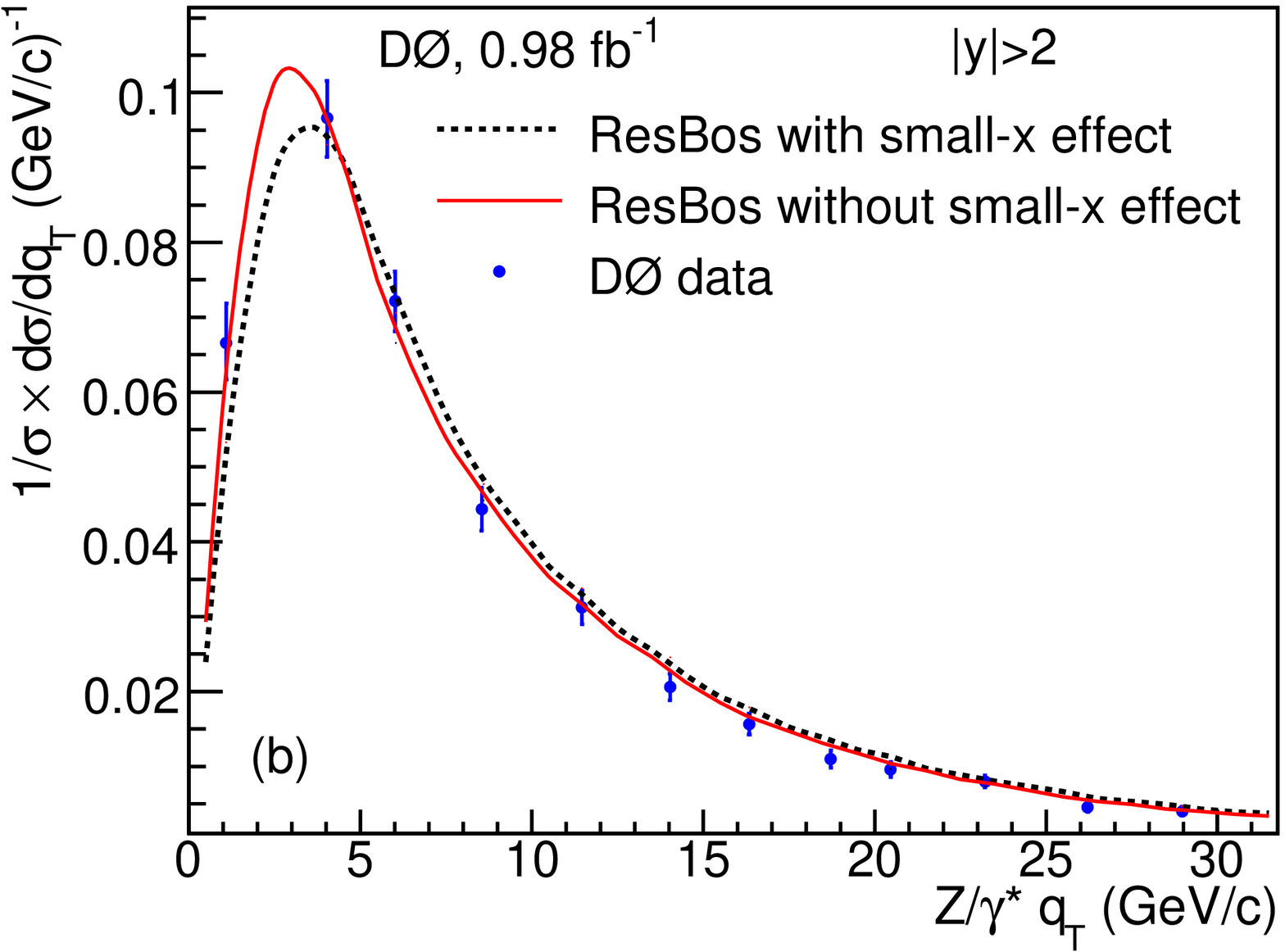}
\caption{Comparison of the theoretical prediction from \textsc{Resbos} for the $Q_T$ spectrum to the experimental data collected by the \dzero collaboration~\cite{D0RunII}, for the inclusive sample (on the left) and for forward rapidities (on the right). This figure is taken from Ref~\cite{D0RunII}.}
\label{fig:resbos}
\end{center}
\end{figure}

The phenomenological tool mostly used by the experimental collaborations is \textsc{Resbos}~\cite{resbos, resbos_comp}. This program uses the resummation formalism developed in Ref.~\cite{CSS} in order to produce NNLL resummed distributions. Moreover, it is matched to fixed-order calculations, in order to provide reliable QCD predictions for all values of $Q_T$. NP effects are also modelled in this code essentially with a Gaussian smearing (BLNY form factor~\cite{resbos_comp}), with parameters fitted to the experimental data. Additionally, phenomenological studies in semi-inclusive deep-inelastic scattering processes suggest an effective dependence of the NP effects on Bjorken $x$. This can be then translated into a small-$x$ contribution to the NP form factor in Drell-Yan processes~\cite{smallxbroad}, usually referred to as small-$x$ broadening. 
In Fig.~\ref{fig:resbos}, we report the comparison between the \textsc{Resbos} theoretical predictions and the data collected by the \dzero collaboration at Tevatron Run II~\cite{D0RunII}. The red line is the theory prediction obtained with the BLNY form factor in \textsc{Resbos}, for the inclusive sample, on the left, and for forward $Z$ boson rapidities, on the right. In the latter case, also the prediction with small-$x$ broadening is plotted. The agreement between theory and experiment is remarkable. Unfortunately, even at large rapidities, where small-$x$ effects are expected to be more pronounced, we cannot discriminate between the two different NP models because of the experimental uncertainty.

\begin{figure}
\begin{center}
\includegraphics[width=0.495 \textwidth]{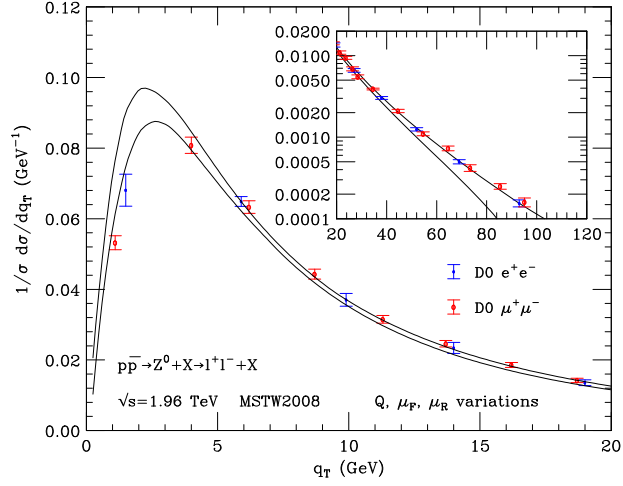}
\caption{Comparison of the theoretical NNLL+NLO prediction of Ref.~\cite{FlorenceDY} for the $Q_T$ spectrum to the experimental data collected by the \dzero collaboration. This figure is taken from Ref.~\cite{FlorenceDY}.}
\label{fig:catani}
\end{center}
\end{figure}

\begin{figure}
\begin{center}
\includegraphics[width=0.495 \textwidth]{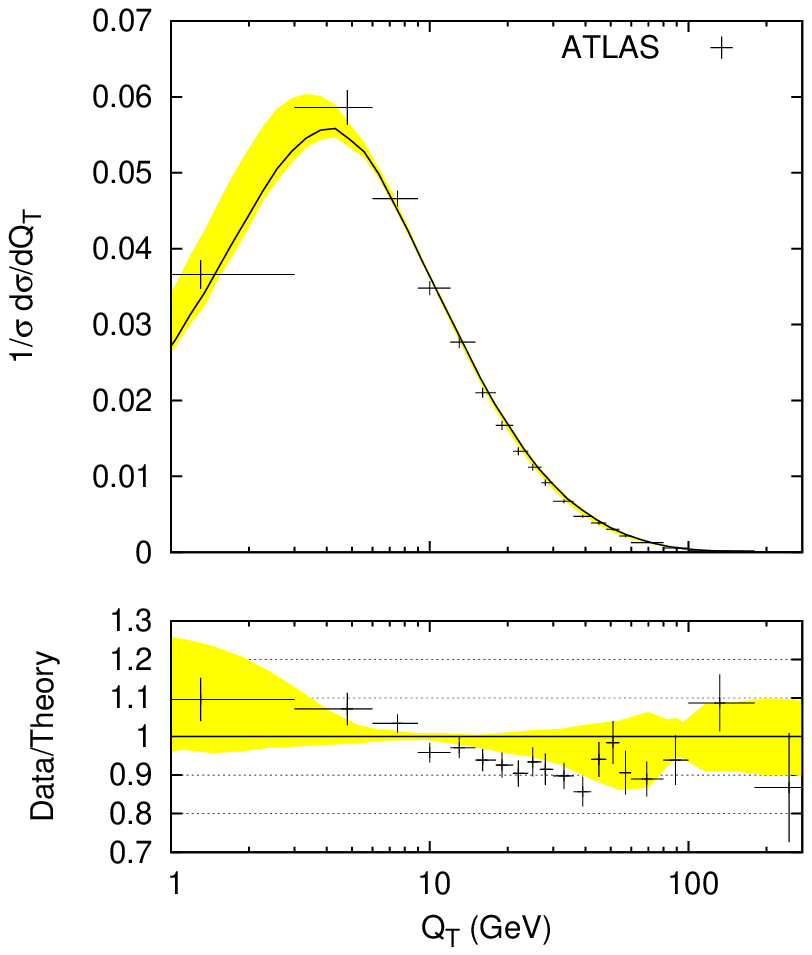}
\includegraphics[width=0.495 \textwidth]{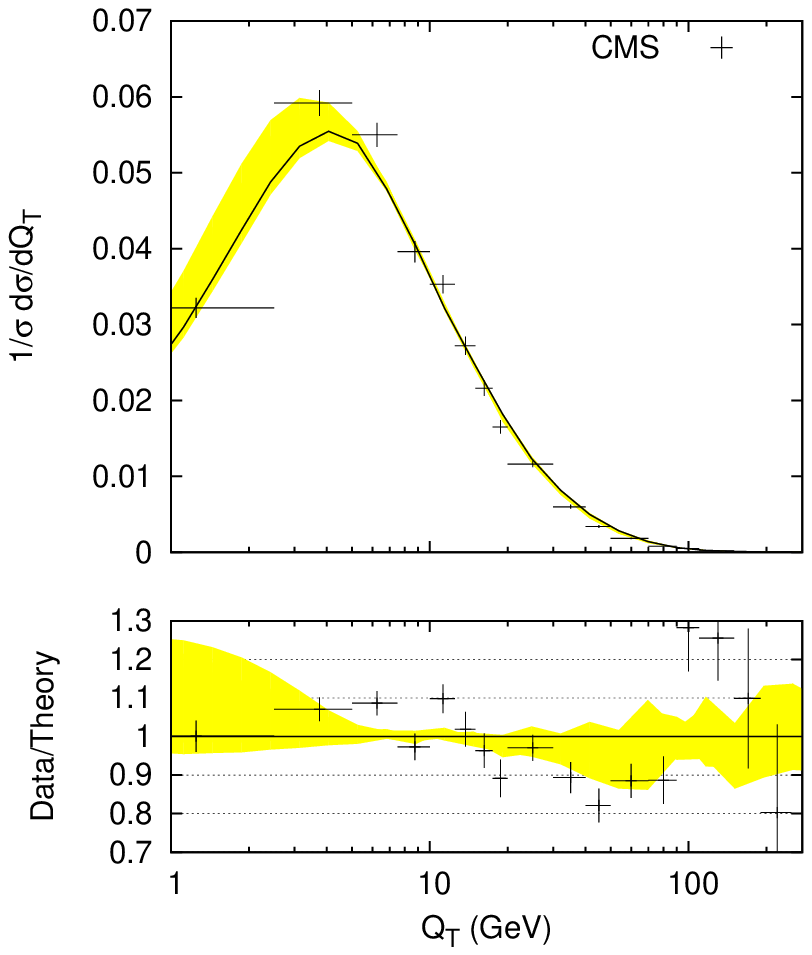}
\caption{Comparison of the theoretical NNLL+NLO prediction~\cite{BDMTlhc} for the $Q_T$ spectrum to the experimental data collected by the ATLAS collaboration~\cite{atlasZpt} (on the left) and by the CMS collaboration~\cite{cmsZpt} (on the right). This figure is taken from Ref.~\cite{BDMTlhc}.}
\label{fig:QTlhc}
\end{center}
\end{figure}

A complementary analysis of the \dzero data was carried out in~\cite{FlorenceDY} and the results are shown in Fig.~\ref{fig:catani}. The theory prediction is a pure perturbative calculation which resums logarithms at NNLL accuracy and it is matched to a NLO calculation. The band represents an estimate of the perturbative uncertainty of the theoretical calculation and, within this error, a good description of the data is found, without any NP correction. The theoretical uncertainty is computed by varying the arbitrary scales present in the calculation. As usual, factorisation and resummation scale variations are considered; however, it turns out that the dominant source of uncertainty at small-$Q_T$ comes from missing higher logarithmic terms, which are estimated by varying the argument of the logarithms we are resumming. 
A pure NNLL+NLO, without any explicit NP effect also gives a good description of LHC data~\cite{BDMTlhc}, as shown in Fig.~\ref{fig:QTlhc}.

\section{Novel variables} \label{novelvariables}
In the previous section we have seen that, despite very accurate calculations, no clear conclusions can be drawn by the comparison of the theory to the $Q_T$ data from the Tevatron experiments. On the theoretical side, we have different ways of estimating the uncertainty as well as different NP models. On the experimental side, we would like to reduce the uncertainty associated with the measurement. To this purpose new variables, labelled $a_T$ and $\phi^*$, were introduced in Refs~\cite{WV, WVBRW}; the $\phi^*$ distribution was measured by the \dzero collaboration~\cite{D0dphi}.
\subsection{Experimental motivations}
The variable $a_T$ is component of $\vec{Q}_T$ orthogonal to the dilepton thrust axis~\cite{WV}:
 \beq \label{aTdef}
 \vec{a}_T = \frac{ \vec{Q}_T \times ( \vec{l}_{T1}- \vec{l}_{T2})}{ |\vec{l}_{T1}- \vec{l}_{T2}|}, 
 \eeq
 where all the above three-vectors are defined by $\vec{v} = (\underline{v},0) $ and $ \underline{l}_{T1,2}$ are the leptons' transverse momenta.  One of the main experimental uncertainties comes the resolution in the measurement of these momenta.  We note that at low $Q_T$ we have:
 \beq \label{aTlowQT}
 a_T\simeq \frac{2 l_{T1}l_{T2}}{l_{T1}+l_{T2}} \sin \Delta \phi,
 \eeq
 where $\Delta \phi$ is the dilepton azimuthal separation, which is close to $\pi$ in the limit we are considering. Thus, at low-$Q_T$, the uncertainty for $a_T$ is reduced with respect to the one for $l_{Ti}$ (and hence $Q_T$) by the presence of the small factor $ \sin \Delta \phi$~\cite{WV}. Moreover, it was noticed that this uncertainty can be further reduced if we rescale $a_T$ by the dilepton invariant mass $M$, essentially because resolution effects partly cancel in the ratio~\cite{WVBRW}. Ideally, we would like to define a variable that depends only on very well-measured angles and, at the same time, maintains a close relation to $Q_T$. 
The observable $\phi^*$ satisfies these properties~\cite{WVBRW}
\begin{equation}
\phi^* = \tan \left (\frac{\pi -\Delta \phi}{2} \right) \sin \theta^*, 
\end{equation}
 where $\theta^*$ is the scattering angle of the dileptons with respect to the beam, in the boosted frame where the leptons are aligned. This angle can be related to the pseudo-rapidities of the leptons:
 $
 \sin \theta^*= 1/\cosh\left(\eta^{(l_1)}-\eta^{(l_2)} \right)$, hence $\phi^*$ is fully determined by angular measurements. 
Moreover, in the low-$Q_T$ limit $\phi^*$ reduces precisely to $a_T/M$.

\subsection{Phenomenology of $\phi^*$} \label{phistar}
We would like to derive a resummed expression for the $\phi^*$ distribution and compare it to the \dzero data~\cite{D0dphi}. We have already seen that, in the relevant $Q_T\to 0$ limit, this variable reduces to $a_T/M$, i.e. one component of the transverse momentum vector. The resummation for $a_T$ was discussed in~\cite{BDDaT} and it can be related to the traditional $Q_T$ resummation described in section~\ref{sec:theory}. 
However, in the present case, we are only interested in one component of $\underline{Q}_T$, rather than its magnitude. This difference results into a cosine rather than the Bessel function $J_0$ in Eq.~(\ref{fact2})
\beq
\label{eq:resummedphistar}
\frac{ {\rm d} \sigma}{ {\rm d} \phi^*} =\frac{1}{\pi} 
\int_0^{\infty} {\rm d} b\, M \,\cos \left(bM \phi^* \right) 
e^{-R(\bar{b},M)} \Sigma \left(x_1,x_2, b,M\right)\,.
\eeq
We first note that the function $R$ which resums the large logarithms in $b$-space is the same as the one obtained for $Q_T$ resummation Eq.~(\ref{Rlambda}). Secondly, the presence of the cosine function has important phenomenological consequences. In fact, the $\phi^*$ distribution does not show the typical Sudakov behaviour as $Q_T$, but rather tends to a constant plateau as $\phi^*\to 0$. We have already discussed that low values of $Q_T$ (and hence also $\phi^*$) can be obtained via Sudakov suppression or kinematical cancellation. In the present case of $\phi^*$ the kinematical cancellation starts to dominate prior to the formation of the Sudakov peak, in contrast to the $Q_T$ case.

\begin{figure}
\begin{center}
\includegraphics[width=0.495 \textwidth]{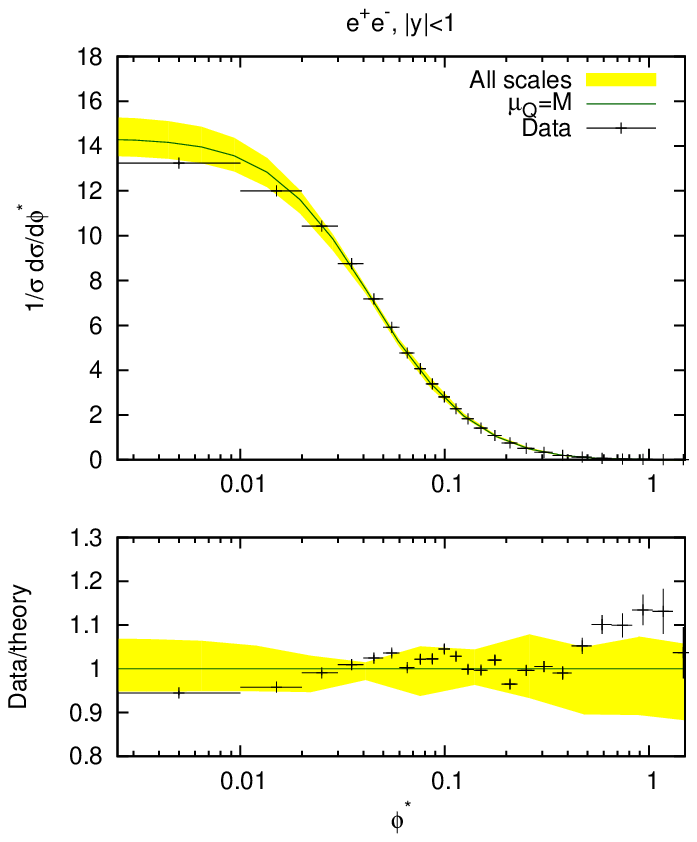}
\includegraphics[width=0.495 \textwidth]{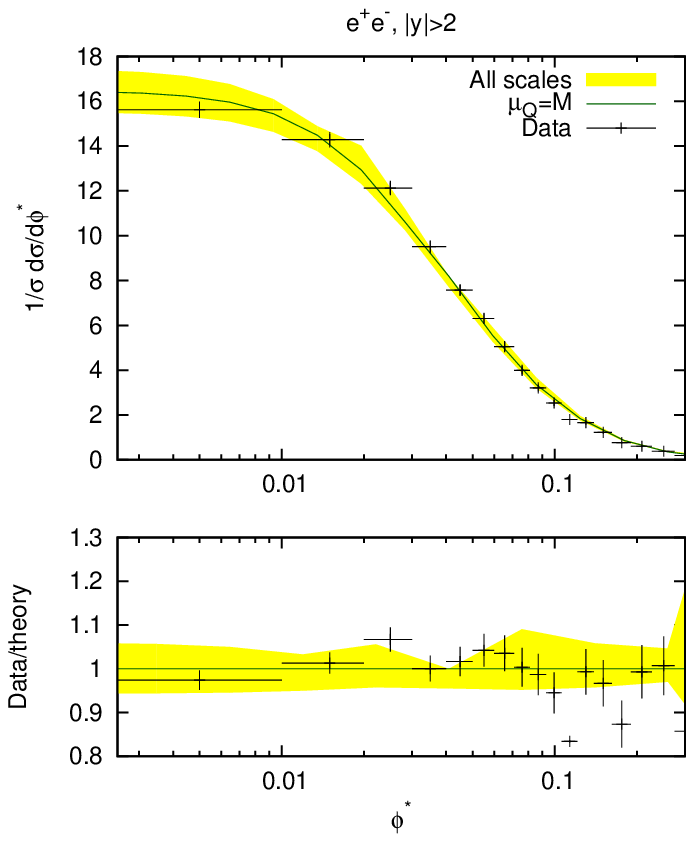}
\caption{Comparison of the theoretical NNLL+NLO prediction~\cite{BDMT} for the $\phi^*$ distribution to the experimental data collected by the \dzero collaboration~\cite{D0dphi}, in different rapidity bins. This figure is taken from Ref.~\cite{BDMT}.}
\label{fig:phistar}
\end{center}
\end{figure}

The \dzero collaboration compared their data to theoretical predictions obtained from the program \textsc{Resbos}. Thanks to the smaller experimental uncertainty the collaboration was able to discriminate between two different NP models, showing in particular that the data at forward rapidities disfavoured small-$x$ broadening, which had not been previously possible due to errors on the $Q_T$ spectrum, even with Tevatron Run-II data~\cite{D0dphi}. 
An independent calculation for the $\phi^*$ distribution was performed in~\cite{BDMT}. NNLL+NLO perturbative predictions were compared to \dzero data with a good agreement in all rapidity regions, once the theoretical uncertainties were faithfully estimated. As an example, we report the comparison between data and theory in Fig.~\ref{fig:phistar} in the case of electrons, for central (left) and forward (right) rapidities.

Predictions for the $\phi^*$ distribution in proton-proton collision at $7$~TeV were first computed in~\cite{BDMTlhc}, where measurements by the LHC collaborations were encouraged. 

\section{Conclusions}
We have reviewed recent phenomenological studies of the transverse momentum distributions of vector bosons produced in hadron-hadron collisions. We have started by discussing the theoretical issues we need to face in order to produce a reliable prediction. The $Q_T$ spectrum can be computed in perturbation theory, but it is affected by large logarithms in the small-$Q_T$ region. These terms can be resummed to all-orders and when the result is matched to fixed-order calculation, we can obtain reliable predictions for all values of $Q_T$.

 We have reported on the comparison between state-of-the-art theoretical calculations for the $Q_T$ spectrum and data coming from the Tevatron and the LHC experiments. In order to reduce the experimental uncertainty, the variable $\phi^*$ was introduced and measured by the \dzero collaboration. We have discussed its relation with $Q_T$ and, consequently, its resummation. The very low experimental systematics makes the variable $\phi^*$ invaluable for precise studies of electro-weak bosons at LHC, with the possibility of probing many subtle QCD effects, such as multi-gluon dynamics and its description within, or beyond, perturbation theory.

\section*{Acknowledgments}
Thanks to Lee Tomlinson for useful comments on the manuscript.
This work is supported by UK's STFC.

\end{document}